%% file: swift_conf_proc_2013.tex
\documentclass[12pt]{article}
\usepackage{graphicx}

\newcommand\pubnumber{Article 42 in eConf C1304143}
\newcommand\pubdate{\today}

\def\napoli{NASA Education and Public Outreach Group\\
Sonoma State University, Rohnert Park, CA 94928}

\def\Title#1{\begin{center} {\Large #1 } \end{center}}
\def\Author#1{\begin{center}{ \sc #1} \end{center}}
\def\Address#1{\begin{center}{ \it #1} \end{center}}

\newcommand\pubblock{\rightline{\begin{tabular}{l} \pubnumber\\
         \pubdate  \end{tabular}}}
\newenvironment{Abstract}{\begin{quotation}  }{\end{quotation}}
\newenvironment{Presented}{\begin{quotation} \begin{center}
             PRESENTED AT\end{center}\bigskip
      \begin{center}\begin{large}}{\end{large}\end{center} \end{quotation}}
\def\Acknowledgements{\bigskip  \bigskip \begin{center} \begin{large}
             \bf ACKNOWLEDGEMENTS \end{large}\end{center}}

\input econfmacros.tex

\begin{document}
\begin{titlepage}
\pubblock

\vfill
\Title{Fourteen Years of Education and Public Outreach for the {\it Swift} Gamma-ray Burst Explorer Mission}
\vfill
\Author{ Lynn Cominsky, Kevin McLin, Aurore Simonnet and the {\it Swift} E/PO Team}
\Address{\napoli}
\vfill
\begin{Abstract}
The Sonoma State University (SSU) Education and Public Outreach (E/PO) group leads the Swift Education and Public Outreach program. For {\it Swift}, we have previously implemented broad efforts that have contributed to NASA's Science Mission Directorate E/PO portfolio across many outcome areas. Our current focus is on highly-leveraged and demonstrably successful activities, including the wide-reaching Astrophysics Educator Ambassador program, and our popular websites: Epo's Chronicles and the Gamma-ray Burst (GRB) Skymap. We also make major contributions working collaboratively through the Astrophysics Science Education and Public Outreach Forum (SEPOF) on activities such as the on-line educator professional development course {\it NASA's Multiwavelength Universe}. Past activities have included the development of many successful education units including the GEMS Invisible Universe guide, the Gamma-ray Burst Educator's guide, and the Newton's Laws Poster set; informal activities including support for the International Year of Astronomy, the development of a toolkit about supernovae for the amateur astronomers in the Night Sky Network, and the {\it Swift} paper instrument and glider models.
\end{Abstract}
\vfill
\begin{Presented}
Huntsville Gamma-ray Burst Symposium\\
Nashville, TN,  April 14--18, 2013
\end{Presented}
\vfill
\end{titlepage}
\def\thefootnote{\fnsymbol{footnote}}
\setcounter{footnote}{0}

\section{Introduction}

We review here the accomplishments of the Sonoma State University (SSU)-led Education \& Public Outreach (E/PO) program for the {\it Swift} mission.  Swift's E/PO programs have been responsive to NASA's increased emphasis on attracting and retaining under-represented STEM (Science, Technology, Engineering and Math) students in the workforce.  We have inspired millions of the public and young people through multi-media presentations, websites and our work with amateur astronomers in the Night Sky Network, bringing exciting {\it Swift} mission science to diverse audiences. We have engaged middle and high-school students and trained thousands of their teachers with our highly popular NASA-approved formal educational products. We have continued this education through after-school programs in partnership with local schools to bring our products and enhanced science education to hundreds of under-served students. We have also engaged  dozens of high school and college students  in authentic NASA research experiences through our Global Telescope Network. All {\it Swift} E/PO program elements are extensively evaluated by external experts at WestEd. Below, we summarize and quantify accomplishments in Formal Education, Informal Education and Public Outreach during the fourteen years of the program.

\section{Formal Education}

{\it Swift}  E/PO promotes STEM careers through the use of NASA data including research experiences for students and teachers (Global Telescope Network), and promotes inquiry into topics that are included in the National Science Education Content Standards A, B, \& D for grades 7-14, including forces and motion, and the relationship between energy and matter. {\it Swift} E/PO links the science objectives of the mission to well-tested, customer-focused and standards-aligned classroom materials (GEMS Invisible Universe guide, Gamma-ray Burst Educator's Guide, Newton's Laws poster set, and the Eyes Through Time curriculum). These materials have been distributed through (Educator Ambassador and {\it NASA's Multiwavelength Universe} on-line) teacher training workshops and through after-school clubs involving under-represented students.

\subsection{Observing with a Robotic Telescope}

The {\it Swift} E/PO program has participated in the development of the Global Telescope Network (GTN). The GTN now consists of 38 partners from small college, high school and amateur observatories from around the world. The goal of the GTN is to partner students and amateur astronomers with members of the science teams from the supporting missions in order to produce publishable multi-wavelength data on targets of scientific interest.  For {\it Swift}, this meant chasing the afterglows which yield information on the counterparts of gamma-ray bursts (GRBs). In support of the GTN, we developed a NASA-approved website ({\tt http://gtn.sonoma.edu} ) with descriptions of hardware, software, observing protocols and target objects. SSU personnel and students observed GRB locations and other targets using our 14-inch GLAST (Gamma-ray Large Area Space Telescope) Optical Robotic Telescope (GORT) at the Pepperwood Preserve and through other telescopes on this network. A major collaborator in the GTN is the Panchromatic Robotic Optical Monitoring and Polarimetry Telescopes (PROMPT) project at the University of North Carolina (headed by Professor Dan Reichart), which runs six telescopes in Chile and also developed the Skynet system for scheduling jobs on multiple telescopes. Using Skynet, we provide GTN access to both northern and southern hemisphere telescopes, using a simple web interface for queuing jobs and retrieving data that works well for high school and college students. Over 200 diverse high school and college students have used the GTN to obtain data in conjunction with SSU E/PO programs. In December 2008, GORT caught its first autonomously-triggered afterglow for GRB 081203A (which was reported in GCN 8617).  This was a very exciting event for all the students who were participating in the GTN activities. In April 2013, GRB 130420A was caught by GORT, and is reported in GCN 14427 and 14445. The telescope was on target about 100 seconds after the trigger, while the burst was still brightening. GORT was still able to detect the GRB 24 hours later at magnitude 21.

\subsection{Curriculum Materials}
The {\it Swift} E/PO team has developed many educational products related to {\it Swift} science that are revised and reprinted as funds permit.  All curricular materials have been approved by NASA Product Review, and all (except the GEMS guide) are available through the {\it Swift} E/PO website {\tt http://swift.sonoma.edu/education/} and NASA Wavelength {\tt (http://nasawavelength.org)}. The products include:

\begin{enumerate}
\item Invisible Universe: From Radio Waves to Gamma Rays: Developed in partnership with the Great Explorations in Math and Science (GEMS) group at the Lawrence Hall of Science © 2004, updated in 2010 and reprinted in 2011. Target audience: grades 6-8. Over 6500 copies of this guide have been sold (including about 2000 distributed through Educator Ambassador workshops). In addition, the Origins missions chose one of the guide's activities to feature on the back of an electromagnetic spectrum wallsheet, which was distributed to over 40,000 members of the National Science Teachers' Association.

\item Gamma-Ray Burst Education Unit: Developed by SSU (and approved by NASA Product Review in 2004), the educator's guide and accompanying wallsheet target grades 9-14. The unit uses GRBs to teach standards-based science concepts, including categorization, plotting, energy transport, the history of science, and science as a human endeavor. More than 3000 have been distributed through educator workshops.

\item Newton's Laws poster set: A set of four educational wallsheets and classroom activities for grades 5-8 that outline Newton's Three Laws of Motion and of Gravitation with ``{\it Swift} connections" to the launch and orbit of {\it Swift}. The posters include activities that are designed to be displayed as a set on middle-school classroom walls, and the pre-activity readings feature {\it Swift}. The poster set was approved by NASA Product Review in 2006, and was most recently reprinted in 2010. More than 7000 of these poster sets have been distributed through educator workshops and by postal mail. In 2012, the posters were downloaded more than 27,000 times.

\begin{figure}[htb]
\centering
\includegraphics[height=1.5in]{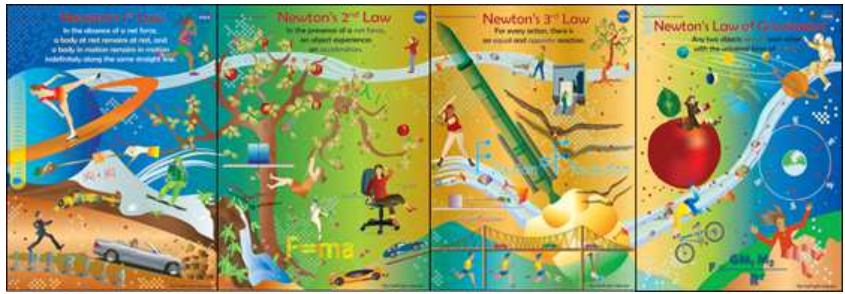}
\caption{Newton's Laws Poster Set}
\label{fig:newton}
\end{figure}

\item {\it Swift} ``Eyes through Time" is a six-unit curriculum for grades 5-8 consisting of an educator's guide accompanied by videos, one for each unit. The curriculum is an outgrowth of Swift-related educational materials developed as part of the ``What's in the News?" program by Penn State Public broadcasting, and scientific oversight was provided by SSU. Approved by NASA Product Review in 2007, the curriculum uses the excitement of gamma-ray burst science to teach standards-aligned topics in science and mathematics that illustrate the scientific process. It is also available through the Teacher's Domain website {\tt http://www.teachersdomain.org/special/kmedia07-ex/scitech.swift/}.

    \end{enumerate}

\subsection{Astrophysics Educator Ambassador Program}
The Astrophysics Educator Ambassadors (EAs) are master teachers who engage other educators in professional development that is directly connected to teaching practices. The EA program currently has 16 educators who annually train thousands of other educators at local, state, regional, and national teachers' meetings, giving SSU-approved workshops across the country. Every other year, they attend a week-long workshop at SSU: the most recent training was held in July 2012, and included a two-day mini-course about Gravitation. Previous mini-courses focused on the dark universe and particle physics. To date, the {\it Swift}-funded EAs, together with SSU E/PO personnel, have directly trained over 24,000 students and teachers through over 360 lectures, mini-courses and workshops. The EA program, as a whole, has directly engaged over 63,500 teachers and students in workshops across North America since 2001. The program is demonstrably successful, as evidenced by an overall evaluation of data from more than 1500 responses (140 workshops) compiled by WestEd. This report indicated that the EA presentations have continued to improve each year, even though ``the means in each year are relatively high for each item (surveyed)." WestEd has also extensively evaluated all five bi-annual EA trainings, many individual workshops conducted by EAs, and some workshops conducted by SSU personnel. For more information about the EA program, to download training materials and videos, or to meet the EAs, see {\tt http://epo.sonoma.edu/ea}.

\subsection{NASA's Multiwavelength Universe}
In 2011, the pilot course {\it NASA's Multiwavelength Universe} was developed, hosted on SSU servers, and Prof. Cominsky served as the Instructor of Record for a diverse group of 25 teachers. The course met the needs of the (predominantly middle-school) teachers by helping them gain content and pedagogical knowledge, as well as experiences with inquiry-based activities and use of the Internet to find NASA resources. The program was evaluated via a pre- and post-test, by collecting feedback from the participants, and also by WestEd. {\it Swift} E/PO continued to support this course during 2012, and plans to continue to support the course through 2016. When the course is not in session, the course materials can be viewed and downloaded at: {\tt http://universe.sonoma.edu/cosmo/course/view.php?id=5}

\subsection{After School Programs}

Beginning in 2004, {\it Swift} E/PO provided partial support (along with {\it Fermi} and {\it XMM-Newton})
for an after-school club that worked with under-represented (primarily
Hispanic) students at Roseland University Prep (9-12) in Santa Rosa, CA. In January
2008, SSU began to support after-school science activities and science and math tutoring
in two different groups at Cali-Calmecac, a "dual immersion" K-8 public charter school
in Windsor, CA. This program was originally a partnership with MESA and then in 2010
became a partnership with the Boys and Girls Club of Northern California. Each semester
we reached approximately 25 students in each group (grades 1-4 and 5-8).  In 2010, we began a program at Roseland Accelerated Middle School (RAMS). The
RAMS after-school program used STEM-focused MESA activities, such as bridge
building, mousetrap cars, and LEGO robotics in order to engage students.

\section{Informal Education and Public Outreach}

{\it Swift} E/PO engages the public in sharing the experience of exploration and discovery through high-leverage multi-media experiences (Black Holes planetarium and PBS NOVA shows, What's in the News? public broadcasts), through popular websites (Gamma-ray Burst Skymap, Epo's Chronicles), social media ({\it e.g.,} Facebook and Twitter), and through activities by amateur astronomers nation-wide (Supernova! toolkit and IYA activities) and inspirational printed materials.

\subsection{Black Hole Shows}

In January 2006 the planetarium show {\it Black Holes: The Other Side of Infinity}, premiered at the Denver Museum of Nature and Science. Jointly funded by NSF and NASA's GLAST (now {\it Fermi}) E/PO program, this full-dome digital show features the {\it Swift} launch, animations of {\it Swift} detecting GRBs and simulations of various aspects of black holes. It is now been shown in over 30 venues worldwide and has reached millions of people. In addition, elements of the planetarium show were used as part of a PBS Nova episode, {\it Monster of the Milky Way}. The episode, which premiered in October of 2006 (and has been rerun several times since then), depicted {\it Swift} detecting a gamma-ray burst and the resulting ground-based follow-up. Initially viewed by over 10 million people and accompanied by additional educational resources, it continues to be available through streaming download from the PBS site: {\tt http://www.pbs.org/wgbh/nova/blackhole}. Prof. Cominsky was the Science Advisor for both shows.

\subsection{What's in the News?}
In partnership with Penn State Public Broadcasting, {\it Swift} E/PO sponsored and provided scientific oversight for the creation of 12 features for What's in the News? -  a news program that was seen by five million middle-school children across the country. It was accompanied by online classroom activities and other supplementary material. In 2004, the funding for the program was eliminated by WPSX, the Penn State public television station. All the {\it Swift} segments are archived at {\tt http://swift.sonoma.edu/program/witn.html}

\subsection{Websites and Social Media}

The {\it Swift} E/PO web site (http://swift.sonoma.edu) is maintained by SSU, but is completely integrated with the mission site (http://swift.gsfc.nasa.gov), providing public access to all the {\it Swift} E/PO materials for the public, students and teachers. Other {\it Swift} E/PO websites include the Gamma-ray Burst (GRB) Skymap site, the GRB Lotto game and Epo's Chronicles, a weekly webcomic.  The GRB Skymap site is aimed at the scientifically attentive public, including many amateur astronomers who ``chase" GRBs and is now supported by both {\it Swift} and {\it Fermi}. The GRB Skymap site includes short writeups of every GRB seen by {\it Swift}, and other satellites, as well as positional information and finding charts. The purpose is to have a complete record of every GRB observed since 2004 (when {\it Swift} was launched, with user-friendly descriptions accessible to the scientifically attentive general public.  To visit the site: {\tt http://grb.sonoma.edu}.

Play the GRB Lotto game, and guess the closest sky location to the next GRB to win: {\tt http://swift.sonoma.edu/grb\_lotto/index.php}.

Each week, SSU staff create, write, and draw the Epo's Chronicles web comic. It is then translated into French, Spanish and Italian. The comic is accompanied by additional background information, definitions of scientific terms, and links to multimedia resources. Monthly readership has doubled in the past year and the strip is now read by over 60,000 unique visitors each year.  Join Alkina and her sentient spaceship Epo, as they travel the galaxy trying to discover their origins and learning about space science: {\tt http://eposchronicles.org}.

{\it Swift} social media pages on Facebook are regularly updated by SSU staff with the latest science news. We also support the {\it Swift} Twitter feed @NASASwift.

\subsection{Amateur Astronomers and the Night Sky Network}

As part of a combined effort with the and {\it Suzaku}, {\it XMM-Newton} and {\it Fermi} E/PO programs, and in partnership with the Astronomical Society of the Pacific (ASP), {\it Swift} E/PO funded and provided scientific oversight of the development of a toolkit for the Night Sky Network of (over 200) amateur astronomy clubs. The kit, entitled {\it Supernova!} was completed during 2008 and, after NASA product approval, went into nationwide distribution. Through 2011, the toolkit reached over 138,000 attendees through 1,284 events.  Of these events, 679 events reported including almost 25,000 minorities and over 39,000 women/girls. We have also worked with a ``High-Energy Network" within the American Association of Variable Star Observers. Many AAVSO members regularly do follow-up observations of GRBs, and some have discovered the visible-light counterparts.

\subsection{International Year of Astronomy (IYA) 2009 Activities }
{\it Swift} E/PO co-sponsored and helped to print a traveling exhibit of 14 images from the collection From Earth to the Universe. Constructed on portable banner stands, this exhibit traveled around the San Francisco Bay Area to 35 venues during 2009 - 2010, with an estimated viewing by over 100,000 participants. SSU personnel accompanied the exhibit to several of the venues, and the banners were designed by SSU illustrator Aurore Simonnet. A larger exhibit with 50 images was also sponsored by the same collaboration, and was displayed at the California Academy of Sciences and the San Jose Tech Museum, and viewed by over 50,000 visitors. During 2009-10, {\it Swift} Educator Ambassadors participated in giving special lectures at (mostly rural) libraries selected to participate in NASA's {\it Visions of the Universe} IYA exhibits. This activity was coordinated by the Astrophysics SEPOF.

\subsection{Printed Materials }
Additional {\it Swift} informal educational products that have been approved by NASA Product Review include the Swift glider model airplane (over 6500 distributed), and the Swift paper model booklet (2500 distributed). Also printed were thousands of {\it Swift} stickers, fact sheets and public brochure.


\section{Evaluation}

Led by Dr. Edward Britton, WestEd conducts independent formative and summative evaluations on a regular basis using professionally accepted qualitative and quantitative assessment tools . Evaluation of the training of teachers, classroom usage and student learning outcomes are the ultimate goals of our assessment process. The results are fed back into the program to improve our effectiveness. WestEd evaluations for Swift include: formative evaluations of the Gamma-ray Burst Educator's Guide, the Newton's Laws poster set, a useability study of the Gamma-ray Burst website, many evaluations of individual workshops by SSU E/PO staff and Educator Ambassadors, longitudinal information on some of the Roseland students who have entered college and progressed into the STEM pipeline, and has provided formative evaluations on our after-school programs which have led to beneficial changes in our work with the students. WestEd has also performed a secondary evaluation of the effectiveness of both offerings of the NASA's Multiwavelength online course, to which Swift was a contributing mission.

\Acknowledgements
We gratefully acknowledge the many SSU staff members and Educator Ambassadors who have supported the {\it Swift} Education and Public Outreach Program since 1999, including Laura Whitlock, Sarah Silva, Phil Plait, Tim Graves, Rob Sparks, Bruce Hemp, Rae McEntyre, Janet Moore, and David Beier.









\end{document}

%% file: econfmacros.tex
\def\beq{\begin{equation}}
\def\eeq#1{\label{#1}\end{equation}}
\def\eeqn{\end{equation}}

\def\beqa{\begin{eqnarray}}
\def\eeqa#1{\label{#1}\end{eqnarray}}
\def\eeqan{\end{eqnarray}}

\let\bar=\overbar

\def\Dslash{\not{\hbox{\kern-4pt $D$}}}
\def\dslash{\not{\hbox{\kern-2pt $\del$}}}

\def\msb{{\bar{\ssstyle M \kern -1pt S}}}

